\documentclass[prd,aps,onecolumn,preprintnumbers,amsmath,amssymb,superscriptaddress]{revtex4}
\pdfoutput=1

\usepackage{amsmath}
\usepackage{graphicx}
\usepackage{dcolumn}
\usepackage{bm}
\usepackage{amssymb}
\usepackage{latexsym}
\usepackage{epstopdf}
\usepackage{threeparttable}
\usepackage{booktabs}
\usepackage{tabularx}
\usepackage{ulem}
\usepackage{float}
\usepackage{CJK}
\usepackage{mathrsfs}
\usepackage{hhline}
\usepackage{multirow}
\usepackage{color}
\usepackage{subfigure}
\usepackage[colorlinks,linkcolor=magenta,anchorcolor=blue,citecolor=blue]{hyperref}

\makeatletter

\newcommand{\Rmnum}[1]{\expandafter\@slowromancap\romannumeral #1@}
\makeatother

\def\be{\begin{equation}}
	\def\ee{\end{equation}}
\def\bea{\begin{eqnarray}}
	\def\eea{\end{eqnarray}}

\begin{document}
\title{New Insights on the Thermodynamic Equilibrium of Black Hole and Thermal Radiation}

\author{Ruifeng Zheng}
\email{zrf2021@mails.ccnu.edu.cn}
\affiliation{Institute of Astrophysics, Central China Normal University, Wuhan 430079, China}

\begin{abstract}
The thermal balance between the black hole and surrounding thermal radiation is an interesting topic. The primordial black hole is a type of black hole formed in the early universe, especially during the thermal radiation period of the early universe. Driven by gravity, thermal radiation will collapse to form a black hole. Because of the complexity of the actual situation, we use the simplest model for simulation under the premise of only considering the thermodynamic properties of the system: in the classic case, an isolated box full of thermal radiation, by adjusting the initial temperature within a reasonable range, the results show that the collapse of part of the thermal radiation to form a black hole and finally achieve thermal equilibrium is an inevitable result of the second law of thermodynamics. We then introduced the quantization of black holes and the number of microscopic states of thermal radiation into the thermal equilibrium system and came to a surprising conclusion: if both the quantization of black holes and the number of microscopic states of thermal radiation are considered at the same time, the fluctuation of thermodynamics is inevitable. When only one of them is considered, the law of conservation of energy and the second law of thermodynamics can be satisfied at the same time, and the fluctuation of thermodynamics can be avoided. This provides a new way of thinking for us to understand the microscopic origin of thermodynamic fluctuations.    
\end{abstract}
\maketitle		
   
 \section{introduction}  
   For an isolated thermodynamic system, the system always tends to thermal equilibrium. This is a corollary of the famous second law of thermodynamics. The universe we live in is like a huge isolated system\cite{Hauret:2017cpn, Saha:2014uwa}, in this isolated system, full of radiation, matter, and black holes. For a huge and complex isolated system like the universe, the study of its thermodynamic properties is extremely complicated. However, we can derive some logical conclusions from the basic theory. For example: Due to the accelerated expansion of the universe, the background temperature of the universe will further drop. The second law of thermodynamics shows that to achieve thermal equilibrium, macroscopically, the temperature of all components in the universe evolve toward the same temperature. Black holes are no exception. Since the mass of the black hole is inversely proportional to the horizon temperature, the future evolution of the universe is more inclined to form massive black holes. 
   \par
   Current theories show that black hole can be formed in two ways: matter shrinks to form black holes and radiation shrinks to form black holes. The black holes formed by radiation collapse in the early universe are called the primordial black hole\cite{Hawking:1975vcx, Franciolini:2021nvv}. However, due to the complex composition of the early universe, the precise composition of the very early universe has not yet been fully clarified. This lack of knowledge may cause us to misunderstand the formation of the primordial black hole. In this paper, we strive to avoid these misunderstandings and explore the process of radiation collapse to form a black hole from a single component model. 
   \par
   The quantization of black holes was first proposed by Bekenstein\cite{Bekenstein:1973ur, Bekenstein:1974jk}. Bekenstein pointed out in his famous paper that the limit of a point particle is illegal in quantum theory, this violates the Heisenberg uncertainty relationship\cite{Bekenstein:1973ur}. Bekenstein believed that particles have a finite proper radius $ b $, and showed that the assimilation of a finite size neutral particle inevitably causes an increase in the horizon area: $ \left( \varDelta A \right) _{\min}=8\pi \mu b $, where $ \mu $ is the rest mass of the particle. the quantization condition of the black-hole surface area can be expressed as $ A_n=\alpha l_{p}^{2}\cdot n\ \ \left( n=1,2,3\cdot \cdot \cdot \right) $\cite{Hod:1998vk}, according to this equation, we can get other quantization conditions\cite{He:2010ct, Hod:2015kpa, Sun:2018muq, Lochan:2015bha}. For a simple thermodynamic system: $ S=k\ln \varOmega\ \ \left( \varOmega=2,3\cdot \cdot \cdot \right) $, $ S $ is the entropy of thermal radiation in the isolated system, $ k $ is Boltzmann constant, $ \varOmega $ is the total number of microscopic states of the system. 
   \par 
   An interesting idea is to link the quantization of black holes with the number of microscopic states of thermal radiation. We introduce it into the study of the thermal balance of the black hole and thermal radiation to obtain the quantum correction of thermal balance conditions. The difference from the classical thermal balance is only in when certain prerequisites are satisfied, the classical thermal balance can be consistent with the quantized thermal balance. A surprising finding is that when the quantization of black holes and the number of microscopic states of thermal radiation are considered at the same time, the thermodynamic fluctuation of the system is inevitable. 
   \par
     The rest of our paper is organized as follows: In Sec. \ref{sec2}, we first analyze the classic second law of thermodynamics. For an isolated box with a certain volume, when the thermal radiation temperature in the box reaches the threshold temperature, the thermal radiation tends to form the black hole. In Sec. \ref{sec3}, we add the analysis of the quantization of black holes and the number of microscopic states of thermal radiation. The calculation shows that the thermodynamic fluctuation is inevitable on the premise of satisfying the law of conservation of energy and the classic second law of thermodynamics. In Sec. \ref{sec4}, we only consider the quantization of black holes, and the results show that when the volume of the box satisfies certain conditions, the law of conservation of energy and the second law of thermodynamics can be satisfied at the same time, and the thermodynamic fluctuation does not exist. In Sec. \ref{sec5}, we only consider the number of microscopic states of thermal radiation, and the results show that when the volume of the box satisfies certain conditions, the law of conservation of energy and the second law of thermodynamics can be satisfied at the same time, and the fluctuation of thermodynamics does not exist. In Sec. \ref{sec6}, we come to our conclusions and discussions.
   	\section{Black hole and thermal radiation}
 \label{sec2}  
\begin{figure}[htbp]\label{FIG1}
	\centering
	\includegraphics[height=2.6cm,width=9.75cm]{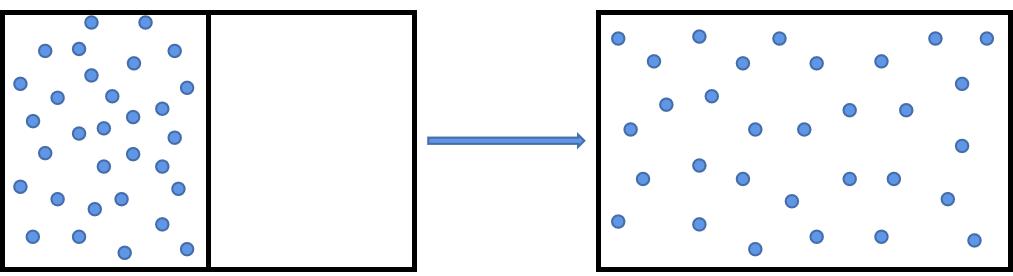}
	\caption{An insulated box is divided into two parts by a heat insulation baffle. The left half is filled with thermal radiation gas molecules. It can be predicted that after some time, the thermal radiation will be evenly distributed in the entire left half, and the left half will maintain thermal equilibrium. Then we pull out the baffle, the thermal radiation from the left half will diffuse freely to the right half, and the final state of the box is shown on the right figure. The thermal radiation will be evenly distributed throughout the box, and the system will reach thermal equilibrium. The entropy of the whole process keeps increasing. }
\end{figure}

\begin{figure}[htbp]\label{2}
	\centering
	\includegraphics[height=2.6cm,width=9.75cm]{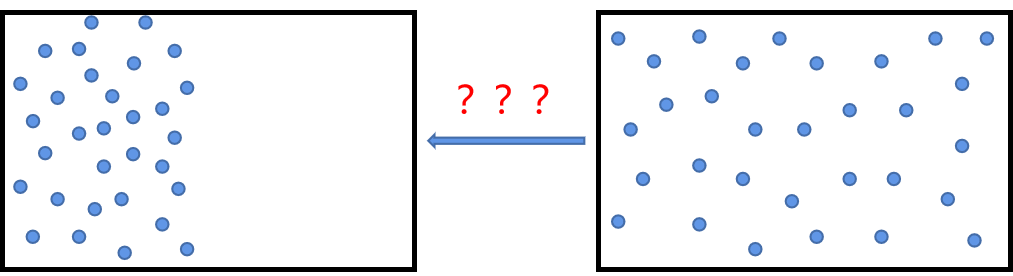}
	\caption{For Fig.\ref{FIG1}, we conclude that after removing the baffle, the thermal radiation gas molecules spontaneously fill the entire box. Interestingly, can this process be reversed? The state of thermal radiation in the right figure will spontaneously return to the state of thermal radiation in the left figure. Through a simple calculation, the entropy of the thermal radiation that fills the entire box is greater than the entropy of the thermal radiation that fills the left half of the box. According to the second law of thermodynamics. It shows that the state of the right figure will not spontaneously change to the state of the left figure. }
\end{figure}

First of all, let’s consider a classic situation, as shown in Fig.\ref{FIG1}. We use a baffle to divide an isolated box into two parts, the left half is full of thermal radiation. After a while, we will find that the thermal radiation will evenly fill the whole in the left half, this is the case allowed by the second law of thermodynamics. At this time, we pull out the middle baffle, the thermal radiation from the left half will be transferred to the right half, and finally, the whole box is evenly filled with thermal radiation. And in this process, we can easily calculate that the entropy of the system is increasing. But the opposite process is shown in Fig.\ref{2}, which is not allowed. This will cause the entropy of the system to decrease. 
\par 
The direction of the passage of time is the direction of increase in entropy, which shows that in an isolated system, the total entropy will not decrease. So this brings us a problem. For an isolated box full of thermal radiation, is the thermodynamic equilibrium of pure radiation the state of maximum entropy of the system? The answer may not be. 
\par
We know that in the early universe, radiation can form the primordial black hole due to gravitational collapse. If the radiation energy of an isolated box is high enough, will the radiation energy collapse to form a black hole? Let’s verify this interesting conjecture. We ignore the influence of the thermodynamic volume of the black hole on the system. 
\begin{figure}[htbp]
	\centering
	\includegraphics[height=2.6cm,width=9.9cm]{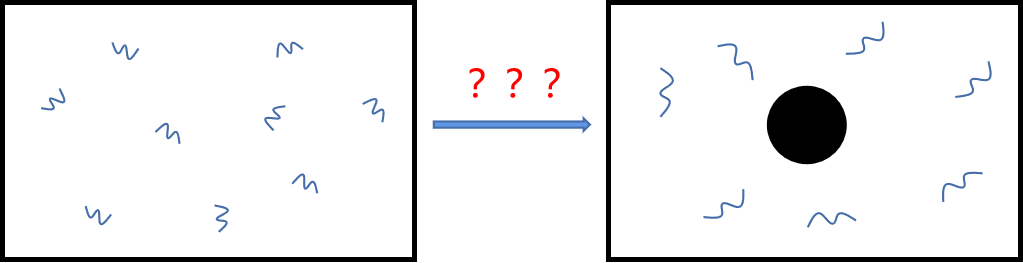}
	\caption{For an isolated box full of thermal radiation, is the thermal equilibrium state in the box really evenly distributed thermal radiation? Is pure uniform thermal radiation stable? Perhaps the answer is not the case. The collapse of thermal radiation to form black hole is a good choice. We do not give the reason for the collapse here, only the inevitability of the collapse into black hole.  }
\end{figure}\label{3}
 \par
When there is only thermal radiation in the box\cite{Gibbons:1976pt}, the radiation energy and radiation entropy are	
\begin{equation}\label{EQ1}	
E_1=aVT_{1}^{4}~,~~~S_1=\frac{4}{3}aVT_{1}^{3}~,
\end{equation}
where $ a=\frac{\pi ^2k^4}{15c^3\textit{$ \hbar $}^3} $, $ T_{1} $ is the initial temperature of the system, $ V $ is the volume of the box. When there are both thermal radiation and the black hole in the box, the total energy, and total entropy are 
\begin{equation}\label{EQ2}
E_2=aVT^{4}+Mc^2~,~~~S_2=\frac{4}{3}aVT^{3}+4\pi k\left( \frac{M}{m_p} \right) ^2~,
\end{equation}
where $ T $ is the temperature of the thermal radiation, $ M $ is the mass of the black hole, $ k $ is Boltzmann constant, $ m_{pl} $ is Planck mass, it can be known from the law of conservation of energy:
\begin{equation}\label{EQ3}
aVT_{1}^{4}=aVT^{4}+Mc^2~. 
\end{equation}
\par
When the thermodynamic system in the box reaches thermal equilibrium, the temperature of thermal radiation is equal to the temperature of the black hole:
\begin{equation}
T_2=T_{BH}~.
\end{equation}
\par
$ T_{2} $ is the temperature of thermal radiation at thermal equilibrium. When the system reaches thermal equilibrium, the entropy of the system reaches its maximum:
\begin{equation}
S_{2}^{'}\left( T_{2} \right) =0~.
\end{equation}
\par
The relationship between the temperature and the mass of the Schwarzschild black hole is
\begin{equation}\label{EQ6}
M=\frac{hc^3}{16\pi ^2kGT_{2}}~.
\end{equation}
\par
In the entire evolution process, substitute Eq.\ref{EQ3} into Eq.\ref{EQ2}, the relationship between the entropy of the system and the temperature can be expressed as (Black hole has formed)
\begin{equation}
S_{2}(T)=\frac{4}{3}aVT^{3}+\frac{4\pi ka^2V^2}{m_{p}^{2}c^4}\left( T_{1}^{4}-T^4 \right) ^2~.
\end{equation}
\par
Since the isolated system evolves in the direction of increasing entropy, when the entropy of the system reaches its maximum value, we have 
\begin{equation}
S_{2}^{'}(T)=\frac{dS}{dT}=4aVT^{2}+\frac{32\pi ka^2V^2}{m_{p}^{2}c^4}\left( T^7-T_{1}^{4}T^3 \right)=0~. 
\end{equation}
\par
Then we get
\begin{equation}
-\frac{m_{p}^{2}c^4}{8\pi kaV}=T_{2}^{5}-T_{1}^{4}T_2~,
\end{equation}
take Eq.\ref{EQ6} into Eq.\ref{EQ3}, then we get
\begin{equation}\label{EQ10}
aVT_{1}^{4}=aVT_{2}^{4}+\frac{hc^5}{16\pi ^2kGT_2}~,
\end{equation}
we set $ \epsilon =\frac{hc^5}{16\pi ^2kG} $, Eq.\ref{EQ10} simplifies to 
\begin{equation}\label{Eq11}
	V(T_{2}^{5}-T_{1}^{4}T_{2})=-\frac{\epsilon}{a}~,
\end{equation}
we substitute the corresponding value:
\begin{eqnarray}\label{EQ12}
	&&k=1.38\times 10^{-23}~J/K~, \nonumber\\
	&&h=6.63\times 10^{-34}~J\cdot s~, \nonumber\\
	&&c=3\times 10^{8}~m/s~, \nonumber\\
	&&G=6.67\times 10^{-11}~N\cdot m^2/kg^2~,
\end{eqnarray}
then Eq.\ref{Eq11} becomes
\begin{equation}
V(T_{2}^{5}-T_{1}^{4}T_{2})=-1.47345\times 10^{55}~m^3\cdot K^5~.
\end{equation}
\par
For a box with a certain volume $ V $, by changing the initial temperature $ T_1 $ of the system, the corresponding thermal equilibrium temperature $ T_2 $ will also change. But it is worth noting that there are two boundary conditions. The first condition is the initial temperature in the box is higher than the temperature at thermal equilibrium. The second condition is the thermodynamic volume of the black hole cannot be larger than the volume of the box, otherwise, it will lead to the collapse of the system, so we have 
\begin{equation}
T_1>T_2~,~\frac{h^3c^3}{384\pi ^5k^3T_{2}^{3}}<V~.
\end{equation}
\par
We define the volume of the box as $ V_{box}=10^{-10}m^3 $, the initial temperature of the system as the only variable. We choose 5 sets of parameters, and the corresponding calculation results are shown in Table.\ref{Table1}.  
\begin{table}
	\centering\caption{$ T_{1} $ is the initial temperature of the system, $ T_{2} $ is the temperature at which the black hole and thermal radiation maintain thermal equilibrium. $ V_{BH} $ is the thermodynamic volume of the black hole, $ S_{1} $ is the initial entropy of the system, $ S_{2} $ is the entropy of the system at thermal equilibrium.}
	\par
	\begin{tabular}{|c|c|c|c|c|c|c|} 
		\hline\rule{0pt}{10pt}
Label &$ V_{box}/m^3 $ & $ T_1/K $	& $ T_2/K $  &  $ V_{BH}/m^3 $ &  $ S_1/(J/K) $&  $ S_2/(J/K) $ \\
		\hline\rule{0pt}{10pt}
		1&$ 10^{-10} $ & $ 4.8\times10^{15} $ & 277.568 & $ 1.19\times 10^{-18} $ & $ 1.10923\times 10^{22} $ & $ 7.19325\times 10^{34} $ \\
		\hline\rule{0pt}{10pt}
		2&$ 10^{-10} $ & $ 4.81\times10^{15} $& 275.267 & $ 1.22\times 10^{-18} $ & $ 1.11618\times 10^{22} $ & $ 7.31401\times 10^{34} $  \\
		\hline\rule{0pt}{10pt}
		3&$ 10^{-10} $ & $ 4.82\times10^{15} $& 272.99 & $ 1.25\times 10^{-18} $ & $ 1.12316\times 10^{22} $ & $ 7.43655\times 10^{34} $ \\
		\hline\rule{0pt}{10pt}
		4&$ 10^{-10} $ & $ 2.4\times10^{16} $ & 0.44411 & $ 2.91\times 10^{-10} $ & $ 1.38654\times 10^{24} $ & $ 2.80986\times 10^{40} $ \\
		\hline\rule{0pt}{10pt}
		5&$ 10^{-10} $ & $ 9\times10^{13} $ & $ 2.24577\times10^{9} $ & $ 2.25\times 10^{-39} $ & $ 7.31183\times10^{16} $ & $ 1.09884\times 10^{21} $\\
		\hline
	\end{tabular}\\
	\label{Table1}
\end{table}
It can be seen from Table.\ref{Table1} that for the first set of data, when the initial temperature is $ 4.8\times10^{15}K $, the pure radiant entropy is $ 1.10923\times 10^{22}J/K $, but if part of the radiation energy is converted into the mass of a black hole, then the temperature of the system is $ 277.568K $ when the thermal equilibrium is reached. At this time, the total entropy of the system is $ 7.19325\times 10^{34}J/K $. Compared with pure radiation, the entropy has increased by about 12 orders of magnitude. We can naturally conclude that according to the second law of thermodynamics when the initial temperature is $ 4.8\times10^{15}K $, the pure thermal radiation system tend to form the black hole.
\par
Compare the first set of data with the second and third sets of data. The results show that as the initial temperature increases, the temperature at the final thermal equilibrium will decrease, which is contrary to our intuition. Since the mass of the Schwarzschild black hole is inversely proportional to the horizon temperature, the increase in the initial temperature will help to form the higher-mass black hole, and the corresponding total entropy will also increase. 
\par
But it is worth noting that the value of the initial temperature is not arbitrary, such as the fourth set of data, the initial temperature is $ 2.4\times 10^{16}K $. Although we can get the temperature at thermal equilibrium and the total entropy of the system when thermal equilibrium is reached, it is worth noting that the volume of the black hole has reached $ 2.91\times10^{-10}m^{3} $ at this time. The thermodynamic volume of the black hole is larger than the volume of the box, which is not allowed. This will lead to the collapse of the entire thermodynamic system, so the initial temperature of $ 2.4\times10^{16}K $ is undesirable. This means that for a box with a fixed volume, there is a maximum temperature limit.
\par  
Since the temperature has an upper limit, is there a lower limit for the temperature? We select the fifth set of data, the initial temperature is $ 9\times10^{13}K $, and we find that the total entropy after reaching the thermal equilibrium is 4 orders of magnitude different from the initial thermal radiation entropy. When the initial temperature is further reduced, we have to face the fact that when the initial temperature is adjusted to a certain value, the thermodynamic system at this time no longer tends to form the black hole, because the total entropy of the system after the formation of the black hole is less than the initial entropy, this violates the second law of thermodynamics. 
\par
As the volume of the box changes, the corresponding upper and lower temperature limits will also change.

	\section{Quantum and microstate }
	 \label{sec3}
According to the famous Boltzmann relation in statistical physics: 
\begin{equation}
S=k\ln \varOmega~,
\end{equation}
combining Eq.\ref{EQ1}, we have
\begin{equation}\label{EQ15}
\frac{4}{3}aVT_{1}^{3}=k\ln \varOmega _1\Rightarrow T_1=\left( \frac{3k\ln \varOmega _1}{4aV} \right) ^{1/3}~,
\end{equation}
\begin{equation}\label{EQ16}
	T_2=\left( \frac{3k\ln \varOmega _2}{4aV} \right) ^{1/3}~,
\end{equation}
here, $ T_{1} $ is the initial temperature of the system, $ T_{2} $ is the temperature when the system reaches thermal equilibrium, $ \varOmega  $ is the number of microscopic states: $ \varOmega=2,3,4,5...m  $.
\par
The quantization condition of the Schwarzschild black-hole surface area is 
\begin{equation}
A_n=\alpha \cdot l_{p}^{2}\cdot n~,
\end{equation}
where $ \alpha=4\ln3 $\cite{Hod:1999oyc}, $ l_{p} $ is Planck length, $ n $ is the quantum number of the Schwarzschild black hole. Correspondingly, we can get the quantization condition of the mass of the Schwarzschild black hole: 
\begin{equation}\label{EQ18}
M_n=\frac{m_p}{4}\cdot \sqrt{\frac{\alpha}{\pi}}\cdot \sqrt{n}~~~~~(n=1,2,3,4,5...)~,
\end{equation}
combining Eq.\ref{EQ3} and Eq.\ref{EQ15}, Eq.\ref{EQ16}, the law of conservation of energy can be written as
\begin{equation}\label{EQ19}
aV\left( \frac{3k\ln \varOmega _1}{4aV} \right) ^{4/3}=aV\left( \frac{3k\ln \varOmega _2}{4aV} \right) ^{4/3}+\frac{m_p}{4}\cdot \sqrt{\frac{\alpha}{\pi}}\cdot \sqrt{n}\cdot c^2~.
\end{equation}
\par
When the system reaches thermal equilibrium, the thermal radiation is equal to the horizon temperature of the black hole, combine Eq.\ref{EQ6}, Eq.\ref{EQ16}, and Eq.\ref{EQ18}, we can get 
\begin{equation}\label{EQ20}
\left( \frac{3k\ln \varOmega _2}{4aV} \right) ^{1/3}=\frac{hc^3}{4\pi ^2kGm_p}\cdot \sqrt{\frac{\pi}{\alpha}}\cdot \frac{1}{\sqrt{n}}\Rightarrow \sqrt{n}=\frac{hc^3}{4\pi ^2kGm_p}\sqrt{\frac{\pi}{\alpha}}\left( \frac{3k\ln \varOmega _2}{4aV} \right) ^{-1/3}~.
\end{equation}
\par
Combining Eq.\ref{EQ12}, then Eq.\ref{EQ20} can be simplified to 
\begin{equation}
\sqrt{n}=7.979\times 10^{33}\cdot V^{1/3}\cdot \left( \ln \varOmega _2 \right) ^{-1/3}~.
\end{equation}
\par
We substitute Eq.\ref{EQ20} into Eq.\ref{EQ19}:

\begin{equation}
aV\left( \frac{3k\ln \varOmega _1}{4aV} \right) ^{4/3}=aV\left( \frac{3k\ln \varOmega _2}{4aV} \right) ^{4/3}+\frac{hc^5}{16\pi ^2kG}\left( \frac{3k\ln \varOmega _2}{4aV} \right) ^{-1/3}~.
\end{equation}
\par
The above formula can be simplified to 

\begin{equation}\label{EQ24}
\left( \ln \varOmega _1 \right) ^{4/3}-\left( \ln \varOmega _2 \right) ^{4/3}=\frac{4 \cdot 3^{-5/3}c^3\pi ^{4/3}}{15^{2/3}hG}\left( \ln \varOmega _2 \right) ^{-1/3}\left( V \right) ^{2/3}~,
\end{equation}
\begin{equation}
\left( \ln \varOmega _1 \right) ^{4/3}-\left( \ln \varOmega _2 \right) ^{4/3}=2.96066\times 10^{68}( \ln \varOmega _2) ^{-1/3}( V ) ^{2/3}~.
\end{equation}

\par
According to the classic second law of thermodynamics, when the system reaches thermal equilibrium, the temperature of the thermal radiation of the box is equal to the horizon temperature of the black hole, which means that Eq.\ref{EQ20} is strictly true. But at the same time, we need to note that $ n $ and $ \varOmega _2 $ of Eq.\ref{EQ20} can only be integers, which means that Eq.\ref{EQ20} can be completely equal only when the volume of the box takes a certain specific value. But under the premise that Eq.\ref{EQ20} is completely equal, we find that Eq.\ref{EQ24} cannot strictly satisfy, this indicates the thermodynamic fluctuation of the system is inevitable. Eq.\ref{EQ20} is an inevitable result obtained under the premise of satisfying the second law of thermodynamics, and Eq.\ref{EQ24} is an inevitable result obtained under the premise of satisfying the law of conservation of energy. The incompatibility of Eq.\ref{EQ20} and Eq.\ref{EQ24} is a result of this paper. This means that for an isolated box in a thermal equilibrium state. Under the premise of considering the quantization of black holes and the number of microscopic states of thermal radiation, as well as strictly satisfy the law of conservation of energy. The thermodynamic fluctuation of the system is inevitable. 
	\section{Quantum number}
	 \label{sec4}
In this part, we only consider the quantization of black holes. In the case of only thermal radiation initially, we once again write the total energy and total entropy of the system: 
	\begin{equation}	
		E_1=aVT_{1}^{4}~,~~~S_1=\frac{4}{3}aVT_{1}^{3}~.
	\end{equation}
\par
After thermal equilibrium, the total energy and total entropy of the system are
	\begin{equation}\label{EQ28}
		E_2=aVT_{2}^{4}+Mc^2~,~~~S_2=\frac{4}{3}aVT_{2}^{3}+4\pi k\left( \frac{M}{m_p} \right) ^2~,
	\end{equation}
take Eq.\ref{EQ18} into Eq.\ref{EQ28}, we get
\begin{equation}
E_2=aVT_{2}^{4}+\frac{m_pc^2}{4}\cdot \sqrt{\frac{\alpha}{\pi}}\cdot \sqrt{n}~.
\end{equation}
\par
The law of conservation of energy can be written as 
\begin{equation}
aVT_{1}^{4}=aVT_{2}^{4}+\frac{m_pc^2}{4}\cdot \sqrt{\frac{\alpha}{\pi}}\cdot \sqrt{n}~.
\end{equation}
\par
When the system reaches thermal equilibrium, the entropy is at its maximum: 
\begin{equation}\label{EQ31}
	-\frac{m_{p}^{2}c^4}{8\pi kaV}=T_{2}^{5}-T_{1}^{4}T_2~,
\end{equation}
where $ T_{2} $ is the temperature at thermal equilibrium, which corresponds to the horizon temperature of the black hole with quantum number $ n $:
\begin{equation}\label{EQ32}
T_2=T_n=\frac{hc^3}{4\pi ^2kGm_p}\cdot \sqrt{\frac{\pi}{\alpha}}\cdot \frac{1}{\sqrt{n}}~,
\end{equation}
take Eq.\ref{EQ32} into Eq.\ref{EQ31}, we get
\begin{equation}\label{EQ33}
-\frac{\gamma}{V_{n}}=\lambda ^5n^{-5/2}-\lambda T_{1}^{4}n^{-1/2}~,
\end{equation}
where $ \gamma=\frac{m_{p}^{2}c^4}{8\pi ka} $, $ \lambda=\frac{hc^3}{4\pi ^2kGm_p}\cdot \sqrt{\frac{\pi}{\alpha}} $. This shows that for a specific initial temperature $ T_{1} $ when the volume $ V_{n} $ of the box takes some specific value, the law of conservation of energy and the second law of thermodynamics can be satisfied at the same time. 
\par
For example, under the premise of determining the quantum number $ n $ of the black hole. Eq.\ref{EQ33} shows that for a given initial temperature $ T_1 $, $ V_n $ can only take a specific value, which can strictly satisfy the law of conservation of energy and the second law of thermodynamics. 
	\section{Number of microscopic states}
	 \label{sec5}
In the previous part, we discussed the requirements of the initial conditions that the system satisfies under the premise of only considering the quantization of black holes. In this part, we only consider the number of microscopic states of thermal radiation, we once again write the Boltzmann relation:   
\begin{equation}
	S=k\ln \varOmega ~.
\end{equation}
\par
The relationship between the thermal radiation temperature $ T $ and the number of microscopic states $ \varOmega $ is 
\begin{equation}\label{EQ37}
	\frac{4}{3}aVT_{1}^{3}=k\ln \varOmega _1\Rightarrow T_1=\left( \frac{3k\ln \varOmega _1}{4aV} \right) ^{1/3}~,
\end{equation}
\begin{equation}\label{EQ38}
	T_2=\left( \frac{3k\ln \varOmega _2}{4aV} \right) ^{1/3}~.
\end{equation}
\par
When the system reaches thermal equilibrium, we have: 
\begin{equation}\label{EQ39}
	-\frac{m_{p}^{2}c^4}{8\pi kaV}=T_{2}^{5}-T_{1}^{4}T_2~,
\end{equation}
we can substitute Eq.\ref{EQ37} and Eq.\ref{EQ38} into Eq.\ref{EQ39} to get 
\begin{equation}
-\eta V_{\varOmega}^{2/3}=\left( \ln \varOmega _2 \right) ^{5/3}-\left( \ln \varOmega _1 \right) ^{4/3}\left( \ln \varOmega _2 \right) ^{1/3}~,
\end{equation}
where $ \eta =\frac{m_{p}^{2}c^4}{8\pi ka}\left( \frac{4a}{3k} \right) ^{5/3} $. $ \varOmega_{1} $ is the number of microscopic states of initial thermal radiation (determined value), $ \varOmega_{2} $ is the number of microscopic states of thermal radiation in thermal equilibrium, $ \varOmega_{1} $ and $ \varOmega_{2} $ are integers. Therefore, when the volume $ V_\varOmega $ of the box takes some specific value, the system strictly satisfies the law of conservation of energy and the second law of thermodynamics. 
	\section{conclusion}
	 \label{sec6}
The formation of the black hole is a hot research direction. This paper takes the collapse of thermal radiation into the black hole as an example to explore the compatibility of the law of conservation of energy and the second law of thermodynamics under the premise of considering the quantization of black holes and the number of microscopic states of thermal radiation. Starting from the classic situation, we use a simple model to discuss the application of the law of conservation of energy and the second law of thermodynamics, and logically give the inevitability of forming the black hole when the radiation temperature of the isolated box reaches a certain value (the inevitability of the increase in entropy of the isolated system). 
\par
In the whole process of the transition from pure radiation to the radiation-black hole and finally to thermal equilibrium. Combine the law of conservation of energy and the second law of thermodynamics, we find that for an isolated box with a certain volume, the thermal radiation temperature has an upper limit and a lower limit. Of course, this is obtained under the premise that the black hole can be formed. When the temperature of thermal radiation is lower than the lower limit, the entire thermodynamic system can only maintain the state of pure thermal radiation. When the threshold temperature is reached (the temperature required to form a black hole), the thermodynamic system tends to form a black hole, because the total entropy of the black hole-thermal radiation thermodynamic system is greater than the total entropy of the pure radiation thermodynamic system. But the temperature cannot rise indefinitely, because the thermodynamic volume of the black hole formed is smaller than the volume of the box, otherwise the entire thermodynamic system will collapse, which is a boundary condition. The second law of thermodynamics indicates that when an isolated system reaches thermal equilibrium, the entropy of the system is maximized. From this, we can get the temperature after the system reaches thermal equilibrium. Under the premise of forming a black hole, by further increasing the initial temperature, we find that the initial temperature is inversely proportional to the temperature at thermal equilibrium, and the initial entropy is proportional to the entropy at thermal equilibrium. This is reasonable, when the initial temperature of the system is higher (the energy is higher), the system is more likely to form a massive black hole.	
\par
According to Boltzmann relation, entropy can be expressed by the number of microscopic states. For a thermodynamic system, the entropy is huge, which indicates that the number of microscopic states is also huge. Through a simple transformation, we can easily get the relationship between the temperature of the thermal radiation system and the number of microscopic states, and the number of microscopic states can only be integers. The quantization of black holes was first proposed by Bekenstein. Hod determined the quantization relationship of the Schwarzschild black-hole surface area in 1998. With the help of the quantization of the black-hole surface area, we can get the quantization of the mass of the black hole. We introduced the quantization of black hole and the number of microscopic states of thermal radiation into the classical thermodynamic analysis of the black hole-thermal radiation system and get an amazing conclusion: under the premise of considering both the quantization of black holes and the number of microscopic states of thermal radiation, thermodynamic fluctuation is inevitable. There is currently no theory that the nature of fluctuation is related to quantization and the number of microscopic states, so we reserve a cautious attitude towards this. 
\par
Since considering both the quantization of black holes and the number of microscopic states of thermal radiation will lead to an amazing conclusion. We separately analyzed the quantization and the number of microscopic states. The results show that under the premise of considering only one of them when the volume of the box takes some special value, the law of conservation of energy and the second law of thermodynamics can be strictly satisfied at the same time (consistent with the classic conclusion). 
	\bibliographystyle{apsrev4-1}
	\bibliography{blackholeandradiation.bib}
\end{document}